\documentstyle[psfig]{cupconf}

\ifoldfss
\else
  \ifnfssone
    \newmathalphabet{\mathit}
      \addtoversion{normal}{\mathit}{cmr}{m}{it}
      \addtoversion{bold}{\mathit}{cmr}{bx}{it}
    \newmathalphabet{\mathcal}
      \addtoversion{normal}{\mathcal}{cmsy}{m}{n}
    \else
    \ifnfsstwo
    \fi
  \fi
\fi

\def\etal{et al.}
\def\kms{km\thinspace s$^{-1}$}

\def\Htwo{{\hbox {H$_2$}}$\;$}

\def\hexnumber#1{\ifcase#1 0\or1\or2\or3\or4\or5\or6\or7\or8\or9\or
 A\or B\or C\or D\or E\or F\fi }

\makeatletter
\ifx\CUP@mtlplain@loaded\undefined
\else
\fi
\makeatother
\makeatletter
\ifx\CUP@mtlplain@loaded\undefined
\else
\fi
\makeatother
\title[H$_{2}$ emission from CRL 618]
{H$_{2}$ emission from CRL 618}

\author[F. Herpin {\it et al.}]
{F. Herpin$^1$,  J. Cernicharo$^1$\\ 
\and \ns A. Heras$^2$}
\affiliation{$^1$Dept F\'{\i}sica Molecular, I.E.M., C.S.I.C, 
Serrano 121, E-28006 Madrid, Spain\\[\affilskip]
$^2$Space Science Dept. of ESA, ESTEC, P.O. Box 299, 
2200 AG Noordwijk, Netherlands}

\begin{document}
\ifnfssone
\else
  \ifnfsstwo
  \else
    \ifoldfss
      \let\mathcal\cal
      \let\mathrm\rm
      \let\mathsf\sf
    \fi
  \fi
\fi

\maketitle

\begin{abstract}
We present a complete study of the \Htwo infrared emission, including 
the pure rotational lines, of 
the proto Planetary 
Nebulae CRL 618 with the ISO SWS. 
A large number of lines are detected. 
The analysis of our observations shows: 
(i) an OTP ratio very different from the classical value 
of 3, probably around 1.76-1.87; (ii) a stratification of the emitting 
region, and more precisely different regions of emission, plausibly located 
in the lobes, in an 
intermediate zone, and close to the torus; (iii) different excitation 
mechanisms, collisions and fluorescence. 
\end{abstract}

\firstsection % if your document starts with a section,
\section{Introduction}

CRL 618 is one of the few clear examples of an AGB
star in the transition phase to the Planetary Nebula stage: a Proto 
Planetary Nebula ({\em PPN}). It has 
a compact HII region created by a hot central C-rich star, and 
is observed as a bipolar nebula at 
optical, radio and infrared wavelengths. 
The expansion velocity of the envelope
is around 20 \kms, but CO observations  
show the presence of a high-velocity outflow with velocities up 
to 300 \kms (Cernicharo \etal 1989). High-velocity emission in 
\Htwo is also detected (Burton \& Geballe 1986).
The high velocity wind and the UV photons from the star 
perturb the circumstellar envelope 
producing shocks and photodissociation regions (PDRs) 
which modify the physical and chemical conditions of the gas
(Cernicharo \etal 1989, and Neri \etal 1992). 
Clumpiness within the visible lobes and low-velocity shocks being the remnant 
of the AGB circumstellar envelope are also proposed by Latter \etal (1992). 
H$_{2}$ excitation mechanisms can be investigated through our 
IR observations. 
We also propose to study the possible 
variations of the OTP ratio from the classical value of 3. 
More details are given in Herpin \etal (2000).

\section{Observations}

Many \Htwo lines were detected (see Fig. 1)
with these ISO Short-Wavelength Spectrometer (SWS, de Grauuw \etal 1996)
observations:
0-0 S(J=0-15), 1-0 Q(J=1-7), 2-1 Q(J=1-7,9,10,13,14,15),
3-2 Q(J=1,3,4,5,7), 4-3 Q(J=2,3,4,7,9,10), and 1-0 O(J=2-7).
The fluxes were obtained assuming the source punctual relative to the SWS
beam. A SWS spectral resolution of $\lambda / \Delta \lambda=2000$ was taken. 

\section{Data analysis}

\subsection{Location of the emission}

Because of the large number of data and thus the large energy covering,
we based our global data analysis on the $0-0$ S transitions. Moreover, 
0-0S(0) to 0-0S(6) emissions are not very sensitive to the extinction and thus are 
very reliable. A brief look on 
the plot of the column densities (derived from the line intensities) 
versus the energies of the upper levels for 
the uncorrected data (no absorption correction, 
and OTP ratio of 3 (Fig.\ref{figuretotal}, left) 
shows
\begin{figure}
  \vspace{0.cm}
  \centerline{\psfig{figure=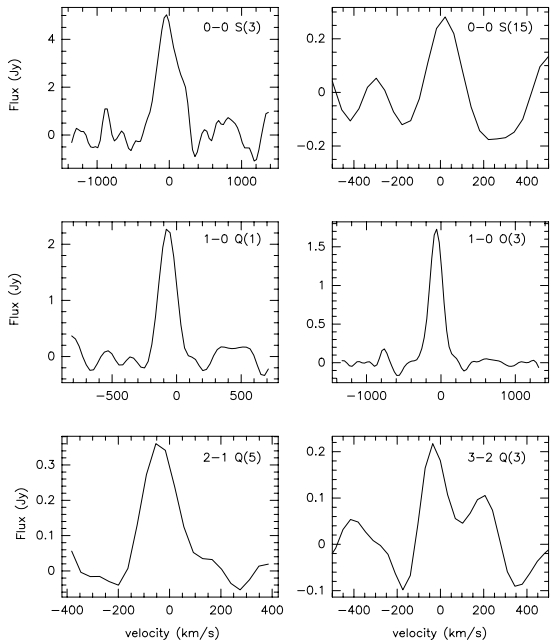,width=6.cm,angle=0}}
  \caption{Examples of observed \Htwo lines centered on the theoretical
wavelength. Fluxes are in Jy. In this Fig., only the $0-0$ S(3) line is
resolved (velocity resolution $\simeq$ 200 \kms).}
  \label{fig1}
\end{figure}
first that an OTP ratio ($\chi$) of 3 seems inappropriate 
as the ortho observations are systematically
underestimated. Moreover, correction of the entire set by the same value of
the OTP ratio seems not possible. 
Thus we introduced different
values for different parts of the plot, i.e., different energies, and thus
probably different regions of excitation in the PPN environment:
$\chi =1., 1.76 $ and 1.87. 
But some discontinuities are still present. 
This is not an effect of variations in $T_{ex}$ because in this case slopes 
would vary. This can only be explained by different 
regions of emission, and consequently can be corrected by different 
absorptions. We thus applied 4 different corrections 
of the absorption: $A= 3.5, 10, 15$ and 20 mag (Latter \etal 1992, 
Thronson 1981). The result is given in the Fig. \ref{figuretotal} 
(right captions). 

\begin{figure} 
  \vspace{0.cm}
  \centerline{\psfig{figure=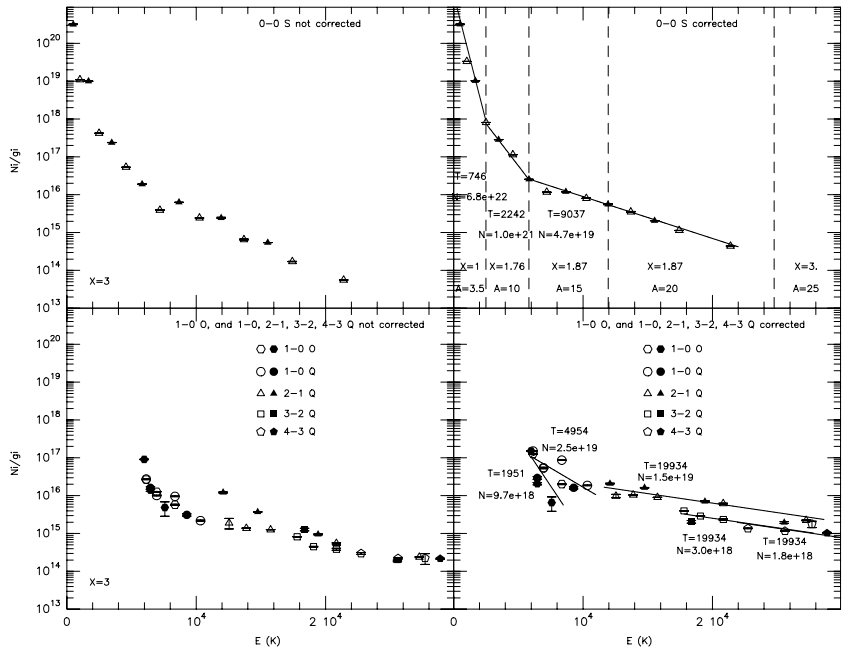,width=10cm,angle=0}}    
  \caption{Plot of the column densities (in cm$^{-2}$) 
    versus the energies of the upper 
    levels for uncorrected data (OTP ratio=3, left captions) and corrected data 
    (right captions). Black and white symbols are respectively for para 
and ortho transitions.}
  \label{figuretotal}
\end{figure} 

This study of the resolution of the observed lines (see Fig. 1) 
shows three different parts of emission. First of all,
the $0-0$ S(0) to S(6) lines are totally resolved (with a SWS resolution of
200 \kms). A broad line, thus resolved, means that the emission arises from a
quite large velocity region, this velocity field being then sufficiently
important to broaden the line. This suggests 
that this part of the \Htwo emission may arise from the lobes, 
where the velocity is larger. 
An intermediate zone of excitation, between lobes and torus, 
appears for the $0-0$ S(7), S(8) 
and S(9) emissions, whose half power line widths are around 200 \kms (third zone 
on the plot) on the limit of resolution. 
Finally, all the other lines have small widths (50-160 \kms), and 
then probably arise from a low-velocity region, certainly the edge of the torus. 
A similar study is done for the emissions involving excited vibrational
levels. The $1-0$ Q lines are unresolved. We think that
these emissions stem from the same intermediate zone as are the $0-0$ S(7)-S(9)
emissions. Moreover the ``intermediate'' slope derived from the column
densities is another indication. The same conclusion applies
for the $1-0$ O lines, even if their origin
is not exactly the same; 
this emission is probably closer to
the lobes, as the derived temperature (cf. Fig.\ref{figuretotal}) 
is similar as what is found
for the $0-0$ S emission in this area. All the
lines involving vibrational levels with $v\geq 2$ are completely unresolved 
and thus arise from a region close to the torus.

\subsection{Excitation mechanisms}

We attempted to fit the data with a very simple model based on a Boltzman
distribution of the populations (cf. Fig. 2). For the $0-0$ S emissions, the fitted temperatures and column
densities are in agreement with the hypothesis of low-energy emission from shocks
in the lobes (low temperatures, and high column
densities, corresponding to
$n(H_{2})\simeq 10^{6}-10^{7}$ cm$^{-3}$), and high-energy
emissions coming from the torus with higher temperature and weaker 
density. For the transitions from excited vibrational levels, 
we admitted that 0-0 S(J), 1-0 Q(J+2) and O(J+4) emissions are produced in the
same region, and have the same extinction and OTP
ratio. The same argument was used for the other excited vibrational
emissions. We obtained different 
fits with slow slopes (temperatures 
are larger) for different $\Delta v$ emissions, which may
suggest fluorescent emission. 

Concerning the $1-0$ emissions, 
we compared our intensities to theoretical results of Sternberg (1989)
and Black (1987). Moreover the temperatures are moderate, 
as the rot-vibrational 
column densities. This suggests the thermal contribution in the excitation. 
A clumpy structure, or one which partly shields
the emitting region from the stellar UV flux may explain a low level 
of fluorescent emission. In thermal sources these rotational and
vibrational temperatures are similar,
while fluorescent sources are characterized by a much higher vibrational
temperature and a much lower rotational temperature (Tanaka \etal 1989). 
That is why
emissions 1-0 O and Q are probably more or less of thermal nature 
($T_{rot} \simeq 2000-5000 $K and $T_{vib} \simeq 1000$ K for $v=1$), 
but a component of fluorescent emission is probably also present. 
The situation seems more complicated for the other vibrational excited transitions.
The only case in Sternberg (1989) and also in Black (1987)
with all these emissions is the radiative fluorescent model. 
This strongly suggests that the emissions 
involving $v \geq 2$ levels are fluorescent. 
\begin{figure} 
  \vspace{0.cm}
  \centerline{\psfig{figure=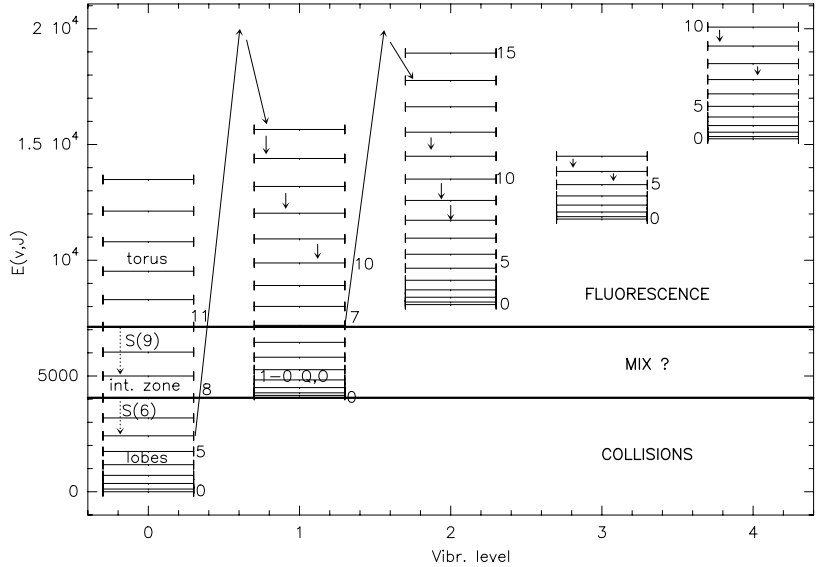,width=10cm,angle=270}}    
  \caption{Energy levels (v,J) of the \Htwo 
    molecule and mechanisms of excitation.}
  \label{energie}
\end{figure} 
We resume on  Fig.\ref{energie} what we deduced 
about the excitation of the \Htwo lines. 

\subsection{The OTP ratio}

We attempt to explain the low OTP ratio by different mechanisms.
Shocks breaking molecules and photodissociation, ionization 
from the central star can produce H, H$^{+}$ and
H$_{3}$$^{+}$\ldots who will react with \Htwo and altere the OTP ratio. 
On the other hand, in a shocked region, 
the shock velocity may be important in the OTP conversion process. 
We think that there is in fact 
no OTP ratio gradient in our data, but large 
uncertainties, and consequently that the OTP ratio is around 1.76-1.87,
the first value of 1 remaining without explanation. We 
thus think that the OTP ratio is fixed near the central star where 
low-velocity shocks occur and where atoms and ions can spoil this ratio. 
Part of this material is swept out into the lobes with this low OTP ratio.

\section{Conclusion}

Shocks in the lobes produce the low-energy 0-0 S emission, while shocks and 
perhaps fluorescence induce emission in a transition area. The high-energy emission 
may be produced by fluorescence 
in the torus. A mix of fluorescence and collisions imay be 
on the origin of the 1-0 O and Q lines, while all the other emissions are 
dominated by fluorescence. Concerning the low OTP ratio, 
probably around 1.76-1.87, we think that it is 
fixed near the central star, the material being after swept out 
into the lobes, with a very non-classical value.

%\begin{acknowledgments}
%We thank spanish DGES and CICYT for funding support
%for this research under grants PB96-0883 and ESP98-1351E.
%\end{acknowledgments}

\end{document}